# Randomized Shellsort: A Simple Data-Oblivious Sorting Algorithm

MICHAEL T. GOODRICH

University of California, Irvine

**Abstract**

In this paper, we describe a **randomized Shellsort** algorithm. This algorithm is a simple, randomized, data-oblivious version of the Shellsort algorithm that always runs in $O(n \log n)$ time and succeeds in sorting any given input permutation with very high probability. Taken together, these properties imply applications in the design of new efficient privacy-preserving computations based on the secure multiparty computation (SMC) paradigm. In addition, by a trivial conversion of this Monte Carlo algorithm to its Las Vegas equivalent, one gets the first version of Shellsort with a running time that is provably $O(n \log n)$ with very high probability.

## 1 Introduction

July 2009 marked the 50th anniversary of the Shellsort algorithm [47]. This well-known sorting algorithm (which should always be capitalized, since it is named after its inventor) is simple to implement. Given a sequence of offset values, $(o_1, o_2, \ldots, o_p)$, with each $o_i < n$, and an unsorted array $A$, whose $n$ elements are indexed from 0 to $n-1$, the Shellsort algorithm (in its traditional form) is as follows:

**for** $i = 1$ to $p$ **do**
   **for** $j = 0$ to $o_i - 1$ **do**
      Sort the subarray of $A$ consisting of indices $j, j+o_i, j+2o_i, \ldots$, e.g., using insertion-sort.

In fact, even this traditional version of Shellsort is actually a family of algorithms, since there are so many different offset sequences. The trick in implementing a traditional version of Shellsort, therefore, is coming up with a good offset sequence. Pratt [38] shows that using a sequence consisting of all products of powers of 2 and 3 less than $n$ results in a worst-case running time of $O(n \log^2 n)$. Several other offset sequences have been studied (e.g., see the excellent survey of Sedgewick [44]), but none beat the asymptotic performance of the Pratt sequence. Moreover, Plaxton and Suel [37] establish a lower bound of $\Omega(n \log^2 n/(\log \log n)^2)$ for the worst-case running time of Shellsort with any input sequence (see also [13]) and Jiang *et al.* [25] establish a lower bound of $\Omega(pn^{1+1/p})$ for the average-case running time of Shellsort. Thus, the only way to achieve an $O(n \log n)$ average-time bound for Shellsort is to use an offset sequence of length $\Theta(\log n)$, and, even then, the problem of proving an $O(n \log n)$ average running-time bound for a version of Shellsort is a long-standing open problem [44].

The approach we take in this paper is to consider a variant of Shellsort where the offset sequence is a fixed sequence, $(o_1, o_2, \ldots, o_p)$, of length $O(\log n)$—indeed, we just use powers of two—but the "jumps" for each offset value in iteration $i$ are determined from a random permutation between two adjacent regions in $A$ of size $o_i$ (starting at indices that are multiples of $o_i$). The standard Shellsort algorithm is equivalent to using the identity permutation between such region pairs, so it is appropriate to consider this to be a randomized variant of Shellsort.

In addition to variations in the offset sequence and how it is used, there are other existing variations to Shellsort, which are based on replacing the insertion-sort in the inner loop with other actions. For instance, Dobosiewicz [14] proposes replacing the insertion-sort with a single linear-time **bubble-sort pass**—doing a left-to-right sequence of compare-exchanges between elements at offset-distances apart—which will work correctly, for example, with the Pratt offset sequence, and which seems to work well in practice for geometric offset sequences with ratios less than 1.33 [14]. Incerpi and Sedgewick [24, 23] study a version of Shellsort that replaces the insertion-sort by a ***shaker*** pass (see also [9, 48]). This is a left-to-right bubble-sort pass followed by a right-to-left bubble-sort pass and it also seems to do better in practice for geometric offset sequences [24]. Yet another modification of Shellsort replaces the insertion-sort with a ***brick*** pass, which is a sequence of odd-even compare-exchanges followed by a sequence of even-odd compare-exchanges [44]. While these variants perform well in practice, we are not aware of any average-case analysis for any of these variants of Shellsort that proves they have an expected running time of $O(n \log n)$. Sanders and Fleischer [41] describe an algorithm they call "randomized Shellsort," which is a data-dependent Shellsort algorithm as in the above pseudo-code description, except that it uses products of random numbers as its offset sequence. They don't prove an $O(n \log n)$ average-time bound for this version, but they do provide some promising empirical data to support an average running time near $O(n \log n)$; see also [33].

## 1.1 Data-Oblivious Sorting

In addition to its simplicity, one of the interesting properties of Shellsort is that many of its variants are ***data-oblivious***. Specifically, if we view compare-exchange operations as a reliable[1] primitive (i.e., as a "black box"), then Shellsort algorithms with bubble-sort passes, shaker passes, brick passes, or any combination of such sequences of data-independent compare-exchange operations, will perform no operations that depend on the relative order of the elements in the input array. Such data-oblivious algorithms have several advantages, as we discuss below.

A data-oblivious algorithm for sorting a set of $n$ items can alternatively be viewed as a ***sorting network*** [27], where the elements in the input array are provided as values given on $n$ input wires and internal gates are compare-exchanges. Ajtai, Komlós, and Szemerédi (AKS) [1] show that one can achieve a sorting network with $O(n \log n)$ compare-exchange gates in the worst case, but their method is quite complicated and has a very large constant factor, even with subsequent improvements [36, 45]. Leighton and Plaxton [29] describe a randomized method for building a data-oblivious sorting network that uses $O(n \log n)$ compare-exchange gates and sorts any given input array with very high probability. Unfortunately, even though the Leighton-Plaxton sorting algorithm is simpler than the AKS sorting network, it is nonetheless considered by some not to be simple in an absolute sense (e.g., see [44]).

One can also simulate other parallel sorting algorithms or network routing methods, but these don't lead to simple time-optimal data-oblivious sequential sorting algorithms. For example, the online routing method of Arora *et al.* [2] is time-optimal but not data-oblivious, as are the PRAM sorting algorithms of Shavit *et al.* [46], Cole [11], Reif [40], and Goodrich and Kosaraju [20]. The shear-sort algorithm of Scherson and Sen [42] is simple and data-oblivious but not time-optimal. The columnsort algorithm of Leighton [28] and the sorting method of Maggs and Vöcking [30] are asymptotically fast, but they both employ the AKS network; hence, they are not simple.

Finally, note that well-known time-optimal sorting algorithms, such as radix-sort, quicksort, heapsort, and mergesort (e.g., see [12, 18, 21, 43]), are not data-oblivious. In addition, well-known data-oblivious sorting algorithms, such as odd-even mergesort and Batcher's bitonic sort (e.g., see [27]), as well as Pratt's version of Shellsort [38], run in $\Theta(n \log^2 n)$ time. Therefore, existing sorting algorithms arguably do not provide a simple data-oblivious sorting algorithm that runs in $O(n \log n)$ time and succeeds with very high probability for any given input permutation.

### 1.1.1 Modern Motivations for Simple Data-Oblivious Sorting

Originally, data-oblivious sorting algorithms were motivated primarily from their ability to be implemented in special-purpose hardware modules [26]. Interestingly, however, there is a new, developing set of applications for data-oblivious sorting algorithms in information security and privacy.

In secure multi-party computation (SMC) protocols (e.g., see [6, 10, 15, 16, 32, 31]), two or more parties separately hold different portions of a set of data values, $\{x_1, x_2, \ldots, x_n\}$, and are interested in computing some function, $f(x_1, x_2, \ldots, x_n)$, on these values. In addition, due to privacy concerns, none of the different parties is willing to reveal the specific values of his or her pieces of data. SMC protocols allow the parties to compute the value of $f$ on their collective input values without revealing any of their specific data values (other than what can inferred from the output function, $f$, itself [19]).

---

[1]We assume throughout this paper that compare-exchange operations always operate correctly; readers interested in fault-tolerant sorting should see, e.g., [3, 8, 17].

One of the main tools for building SMC protocols is to encode the function $f$ as a circuit and then simulate an evaluation of this circuit using digitally-masked values, as in the Fairplay system [6, 31]. By then unmasking only the output value(s), the different parties can learn the value of $f$ without revealing any of their own data values. Unfortunately, from a practical standpoint, SMC systems like Fairplay suffer from a major efficiency problem, since encoding entire computations as circuits can involve significant blow-ups in space (and simulation time). These blow-ups can be managed more efficiently, however, by using data-oblivious algorithms to drive SMC computations where only the primitive operations (such as MIN, MAX, AND, ADD, or compare-exchange) are implemented as simulated circuits. That is, each time such an operation is encountered in such a computation, the parties perform an SMC computation to compute its masked value, with the rest of the steps of the algorithm performed in an oblivious way. Thus, for a problem like sorting, which in turn can be used to generate random permutations, in a privacy-preserving way, being able to implement the high-level logic in a data-oblivious manner implies that simulating only the low-level primitives using SMC protocols will reveal no additional information about the input values. This zero-additional-knowledge condition follows from the fact that data-oblivious algorithms use their low-level primitive operations in ways that don't depend on input data values. Therefore, we would like to have a simple data-oblivious sorting algorithm, so as to drive efficient SMC protocols that use sorting as a subroutine.

## 1.2 Our Results

In this paper, we present a simple, data-oblivious randomized version of Shellsort, which always runs in $O(n \log n)$ time and sorts with very high probability. In particular, the probability that it fails to sort any given input permutation will be shown to be at most $1/n^b$, for constant $b \geq 1$, which is the standard for "very high probability" (v.h.p.) that we use throughout this paper.

Although this algorithm is quite simple, our analysis that it succeeds with very high probability is not. Our proof of probabilistic correctness uses a number of different techniques, including iterated Chernoff bounds, the method of bounded average differences for Doob martingales, and a probabilistic version of the zero-one principle. Our analysis also depends on insights into how this randomized Shellsort method brings an input permutation into sorted order, including a characterization of the sortedness of the sequence in terms of "zones of order." We bound the degree of zero-one unsortedness, or ***dirtiness***, using three probabilistic lemmas and an inductive argument showing that the dirtiness distribution during the execution of our randomized Shellsort algorithm has exponential tails with polylogarithmic dirtiness at their ends, with very high probability (w.v.h.p.). We establish the necessary claims by showing that the region compare-exchange operation simultaneously provides three different kinds of near-sortedness, which we refer to as a "leveraged-splitters." We show that, as the algorithm progresses, these leveraged-splitters cause the dirtiness of the central region, where zeroes and ones meet, to become progressively cleaner, while the rest of the array remains very clean, so that, in the end, the array becomes sorted, w.v.h.p.

In addition to this theoretical analysis, we also provide a Java implementation of our algorithm, together with some experimental results.

As a data-oblivious algorithm, our randomized Shellsort method is a Monte Carlo algorithm (e.g., see [34, 35]), in that it always runs in the same amount of time but can sometimes fail to sort. It can easily be converted into a data-dependent Las Vegas algorithm, however, which always succeeds but has a randomized running time, by testing if its output is sorted and repeating the algorithm if it is not. Such a data-dependent version of randomized Shellsort would run in $O(n \log n)$ time with very high probability; hence, it would provide the first version of Shellsort that provably runs in $O(n \log n)$ time with very high probability.

## 2 Randomized Shellsort

In this section, we describe our randomized Shellsort algorithm. As we show in the sections that follow, this algorithm always runs in $O(n \log n)$ time and is highly likely to succeed in sorting any given input permutation.

Suppose that we are given an $n$-element array, $A$, that we wish to sort, where we assume, without loss of generality, that $n$ is a power of 2. Our randomized Shellsort algorithm uses a geometrically decreasing sequence of offsets, $O = \{n/2, n/4, n/8, \ldots, 1\}$. For each offset, $o \in O$, we number consecutive regions in $A$ of length $o$, as 0, 1, 2, etc., with each starting at an index that is a multiple of $o$, so that region 0 is $A[0\,..\,o-1]$, region 1 is $A[o\,..\,2o-1]$, and so on. We compare pairs of regions according to a schedule that first involves comparing regions by a **shaker pass**—an increasing sequence of adjacent-region comparisons followed by a decreasing sequence of adjacent-region comparisons. We then perform an **extended brick pass**, where we compare regions that are 3 offsets apart, then regions 2 offsets apart, and finally those that are odd-even adjacent and then those that are even-odd adjacent. We refer to this entire schedule as a **shaker-brick** pass, since it consists of a shaker pass followed by a type of brick pass.

## 2.1 Region Compare-Exchange Operations

Each time we compare two regions, say $A_1$ and $A_2$, of $A$, of size $o \in O$ each, we form $c$ independent random matchings of the elements in $A_1$ and $A_2$, for a constant $c \geq 1$, which is determined in the analysis. For each such matching, we perform a compare-exchange operation between each pair of elements in $A_1$ and $A_2$ that are matched, in an iterative fashion through the $c$ matchings. We refer to this collective set of compare-exchanges as a **region compare-exchange**. (See Figure 1.)

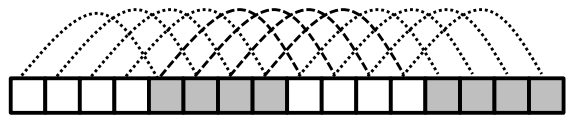

(a)

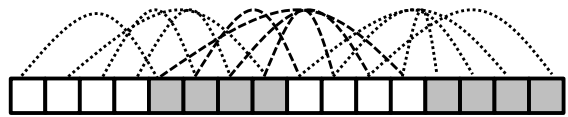

(b)

Figure 1: The region compare-exchange operation. Connections are shown between pairs of regions colored white and their neighboring regions colored gray, under (a) the identity permutation and (b) a random permutation for each pair.

**for** $o = n/2, n/2^2, n/2^3, \ldots, 1$ **do**
    Let $A_i$ denote subarray $A[io\,..\,io+o-1]$, **for** $i = 0, 1, 2, \ldots, n/o - 1$.
    **do** a shaker pass:
        Region compare-exchange $A_i$ and $A_{i+1}$, **for** $i = 0, 1, 2, \ldots, n/o - 2$.
        Region compare-exchange $A_{i+1}$ and $A_i$, **for** $i = n/o - 2, \ldots, 2, 1, 0$.
    **do** an extended brick pass:
        Region compare-exchange $A_i$ and $A_{i+3}$, **for** $i = 0, 1, 2, \ldots, n/o - 4$.
        Region compare-exchange $A_i$ and $A_{i+2}$, **for** $i = 0, 1, 2, \ldots, n/o - 3$.
        Region compare-exchange $A_i$ and $A_{i+1}$, **for** even $i = 0, 1, 2, \ldots, n/o - 2$.
        Region compare-exchange $A_i$ and $A_{i+1}$, **for** odd $i = 0, 1, 2, \ldots, n/o - 2$.

Figure 2: A Pseudo-code description of our randomized Shellsort algorithm.

## 2.2 The Core Algorithm

A pseudo-code description of our randomized Shellsort algorithm, which assumes $n$ is a power of two, is as shown in Figure 2. (We also provide a complete Java implementation in Figure 5, in Section 5.)

Clearly, the description of our randomized Shellsort algorithm shows that it runs in $O(n \log n)$ time, since we perform $O(n)$ compare-exchange operations in each of $\log n$ iterations.

## 2.3 Adding a Cleanup Phase

Even though the above randomized Shellsort algorithm works, as is, in practice (e.g., see Section 5), we make a minor addition to the core algorithm here for the sake of proving a high-probability bound. In particular, we add a cleanup postprocessing phase at the end of the core algorithm that takes care of any stray elements that are out of place, provided there are not too many such elements. This modification is probably an artifact of our analysis, not the algorithm itself, but it is nevertheless helpful in proving a high-probability bound.

Define an $n$-element array, $A$, to be $m$-***near-sorted*** if all but $m$ of the $n$ elements in $A$ are in sorted order. A $p$-***sorter*** [4, 5] is a deterministic sorting algorithm that can sort a subarray of size $p$ as an atomic action. Suppose $\mathcal{S}$ is a data-oblivious (deterministic) $2m$-sorter that runs in $T(m)$ time. Define an $\mathcal{S}$-***shaker pass*** over $A$ to consist of a use of $\mathcal{S}$ at positions that are multiples of $m$ going up $A$ and then down. That is, an $\mathcal{S}$-shaker pass is defined as follows:

**for** $i = 0$ to $n - 2m$ incrementing by steps of $m$ **do**
  Use $\mathcal{S}$ to sort $A[i \mathbin{..} i + 2m - 1]$.
**for** $i = n - 2m$ to $0$ decrementing by steps of $m$ **do**
  Use $\mathcal{S}$ to sort $A[i \mathbin{..} i + 2m - 1]$.

To show that this method sorts an $m$-near-sorted array $A$, we make use of the ***zero-one principle for sorting networks*** (which also applies to data-oblivious sorting algorithms):

**Theorem 2.1 ((Knuth [26]))** *A sorting network (or data-oblivious sorting algorithm) correctly sorts all sequences of arbitrary inputs if and only if it correctly sorts all sequences of 0-1 inputs.*

The main idea behind this principle is that it allows us to reduce each case of distinguishing the $k$ largest elements and the $n - k$ smallest elements to an instance having $k$ ones and $n - k$ zeroes. This allows us to easily prove the following:

**Lemma 2.2** *Given an $m$-near-sorted array $A$ of size $n$, and a $2m$-sorter $\mathcal{S}$, running in $T(m)$ time, a $\mathcal{S}$-shaker pass over $A$ will sort $A$ in $O(T(m)n/m)$ time.*

**Proof:** Suppose $A$ is an $m$-near-sorted binary array, consisting of $k$ ones and $n - k$ zeroes. Thus, there are at most $m$ ones below position $n - k$ in $A$ and at most $m$ zeroes after this position in $A$. Since it sorts subarrays of size $2m$ in an overlapping way, the forward loop in an $\mathcal{S}$-shaker pass will move up all the lower-order ones so that there are no ones before position $n - k - m'$, where $m'$ is the number of high-order zeroes. Thus, since $m' \leq m$, the backward loop in an $\mathcal{S}$-shaker pass will move down all high-order zeroes so that there are no zeroes after position $n - k$. ∎

We show below that the randomized Shellsort, as described in Section 2, will $\alpha \operatorname{polylog}(n)$-near-sort an input array $A$, with very high probability, for some constant $\alpha > 0$. We can then use Pratt's version [38] of (deterministic) Shellsort as a $2\alpha \operatorname{polylog}(n)$-sorter, $\mathcal{S}$, in a $\mathcal{S}$-shaker postprocessing pass over $A$, which will run in $O(n(\log \log n)^2)$ time and (by Lemma 2.2) will complete the sorting of $A$. Note, in addition, that since we are using a Shellsort implementation in an $\mathcal{S}$-shaker (Shellsort-type) pass, adding this postprocessing phase to our randomized Shellsort algorithm keeps the entire algorithm being a data-oblivious variant of the Shellsort algorithm.

# 3 Analyzing the Region Compare-Exchange Operations

Let us now turn to the analysis of the ways in which region compare-exchange operations bring two regions in our size-$n$ input array closer to a near-sorted order. We begin with a definition.

## 3.1 Leveraged Splitters

Ajtai, Komlós, and Szemerédi [1] define a $\lambda$-*halver* of a sequence of $2N$ elements to be an operation that, for any $k \leq N$, results in a sequence so that at most $\lambda k$ of the largest $k$ elements from the sequence are in the first $k$ positions, and at most $\lambda k$ of the smallest $k$ elements are in the last $k$ positions. We define a related notion of a $(\mu, \alpha, \beta)$-*leveraged-splitter* to be an operation such that, for the $k \leq (1-\epsilon)N$ largest (resp., smallest) elements, where $0 \leq \epsilon < 1$, the operation returns a sequence with at most

$$\max\{\alpha(1-\epsilon)^\mu N,\ \beta\}$$

of the $k$ largest (smallest) elements on the left (right) half. Thus, a $\lambda$-halver is automatically a $(1, \lambda, 0)$-leveraged-splitter, but the reverse implication is not necessarily true. The primary advantage of the leveraged-splitter concept is that it captures the way that $c$ random matchings with compare-exchanges has a modest impact with respect to a roughly equal number of largest and smallest elements, but they have a geometric impact with respect to an imbalanced number of largest and smallest elements. We show below that a region compare-exchange operation consisting of at least an appropriate constant number of random matchings is, with very high probability, a $(\alpha, \beta, \mu)$-leveraged-splitter for each of the following sets of parameters:

$$\mu = c+1, \quad \alpha = 1/2, \quad \text{and} \quad \beta = 0,$$
$$\mu = c+1, \quad \alpha = (2e)^c, \quad \text{and} \quad \beta = 4e\log n,$$
$$\mu = 0, \quad \alpha = 1/6, \quad \text{and} \quad \beta = 0,$$
$$\mu = c+1, \quad \alpha = 1/2, \quad \text{and} \quad \beta = N/20,$$

The fact that the single region compare-exchange operation is a $(\mu, \alpha, \beta)$-leveraged-splitter for each of these different sets of parameters, $\mu$, $\alpha$, and $\beta$, allows us to reason about vastly divergent degrees of sortedness of the different areas in our array as the algorithm progresses. For instance, we use the following lemma to reason about regions whose sortedness we wish to characterize in terms of a roughly equal numbers of smallest and largest elements.

**Lemma 3.1** *Suppose a $(0, \lambda, 0)$-leveraged-splitter is applied to a sequence of $2N$ elements, and let $(1-\epsilon)N \leq k \leq (1+\epsilon)N$ and $l = 2N - k$, where $0 < \lambda < 1$ and $0 \leq \epsilon < 1$. Then at most $(\lambda + \epsilon)N$ of the $k$ largest elements end up in the left half of the sequence and at most $(\lambda + \epsilon)N$ of the $l$ smallest elements end up in the right half of the sequence.*

**Proof:** Let us consider the $k$ largest elements, such that $(1-\epsilon)N \leq k \leq N$. After applying a $(0, \lambda, 0)$-leveraged-splitter, there are at at most $\lambda N$ of the $k$ largest elements on the left half of the sequence (under this assumption about $k$). Then there are at least $N - \lambda N$ of the $l$ smallest elements on the left left; hence, at most $l - (N - \lambda N)$ of the $l$ smallest elements on the right half. Therefore, there are at most $N + \epsilon N - (N - \lambda N) = (\lambda + \epsilon)N$ of the $l$ smallest elements on the right half. A similar argument applies to the case when $N \leq k \leq (1+\epsilon)N$ to establish an upper bound of at most $(\lambda + \epsilon)N$ of the $k$ smallest elements on the left half. ∎

Let us know turn to the proofs that region compare-exchange operations are $(\mu, \alpha, \beta)$-leveraged-splitters, for each of the sets of parameters listed above.

## 3.2 The $(c+1, 1/2, 0)$-Leveraged-Splitter Property

We begin with the $(c+1, 1/2, 0)$-leveraged-splitter property. So suppose $A_1$ and $A_2$ are two regions of size $N$ each that are being processed in a region compare-exchange operation consisting of $c$ random matchings.

We wish to show that, for the $k \leq (1-\epsilon)N$, largest (resp., smallest) elements, this operation returns a sequence such that there are at most
$$\frac{(1-\epsilon)^{c+1}}{2}N$$
of the $k$ largest (smallest) elements on the left (right) half. Without loss of generality, let us focus on the $k$ largest elements. Furthermore, let us focus on the case where largest $k$ elements are all ones and the $2N - k$ smallest elements are all zeroes, since a region compare-exchange operation is oblivious. That is, if a region compare-exchange operation is a $(c+1, 1/2, 0)$-leveraged-slitter in the zero-one case, then it is a $(c+1, 1/2, 0)$-leveraged-slitter in the general case as well.

**Lemma 3.2** *Suppose we are given two regions, $A_1$ and $A_2$, each of size $N$, and let $k = k_1 + k_2$, where $k_1$ (resp., $k_2$) is the number of ones in $A_1$ (resp., $A_2$). Let $k_1^{(1)}$ be the number of ones in $A_1$ after a single region compare-exchange matching. Then*
$$E(k_1^{(1)}) = k_1\left(\frac{k_2}{N}\right).$$

**Proof:** In order for a one to remain on the left side after a region compare-exchange matching, it must be matched with a one on the right side. The probability that a one on the left is matched with a one on the right is $k_2/N$. ∎

We use the above lemma in the proof of the following, which applies to the case when $\epsilon$ is relatively large (that is, when $(1-\epsilon)$ is relatively small).

**Lemma 3.3 ((Fast-Depletion Lemma))** *Given two binary regions, $A_1$ and $A_2$, each of size $N$, let $k = k_1 + k_2$, where $k_1$ and $k_2$ are the respective number of ones in $A_1$ and $A_2$, and suppose $k \leq (1-\epsilon)N$, for $1/4 \leq \epsilon < 1$. Let $k_1^{(c)}$ be the number of ones in $A_1$ after $c$ random matchings (with compare-exchanges) in a region compare-exchange operation. Then*
$$\Pr\left(k_1^{(2c-1)} > \frac{(1-\epsilon)^{c+1}N}{2}\right) \leq (2c-1)e^{-(1-\epsilon)^{c+1}N/2^{10}}.$$

**Proof:** The proof is by induction on the number, $c$, of random matchings. By a theorem of Hoeffding [22], the expected value of any convex function of the size of such a sample is bounded by the expected value of that function applied to the size of a similar sample with replacement. Thus, we can apply a Chernoff bound (e.g., see [34, 35]) to this single random matching and pairwise set of compare-exchange operations. Note that, for the base case,
$$E(k_1^{(1)}) \leq k_1\left((1-\epsilon) - k_1/N\right),$$
which is maximized for $k_1 = (1-\epsilon)N/2$. Thus,
$$E(k_1^{(1)}) \leq \frac{(1-\epsilon)^2 N}{4}.$$
Therefore, by a well-known Chernoff bound (e.g., see [34, 35]),
$$\Pr\left(k_1^{(1)} > \frac{(1-\epsilon)^2 N}{2}\right) \leq \left(\frac{e}{4}\right)^{(1-\epsilon)^2 N/4} \leq e^{-(1-\epsilon)^2 N/2^4},$$
which establishes the base case.

For the inductive case, $c \geq 2$, let us assume inductively that
$$\Pr\left(k_1^{(2c-3)} > \frac{(1-\epsilon)^c N}{2}\right) \leq (2c-3)e^{-(1-\epsilon)^c N/2^{10}}.$$
Recall that
$$E(k_1^{(2c-2)}) \leq k_1^{(2c-3)}\left((1-\epsilon) - k_1^{(2c-3)}/N\right),$$

which is maximized, in this case, for $k_1^{(2c-3)} = (1-\epsilon)^c N/2$, since $x(1-x)$ is a monotonic function for $x \in [0, 1/2]$. Thus,

$$E(k_1^{(2c-2)}) \leq \frac{(1-\epsilon)^{c+1}N}{2} \leq \left(\frac{3}{4}\right)\frac{(1-\epsilon)^c N}{2},$$

since $(1-\epsilon) \leq 3/4$. Therefore, by a well-known Chernoff bound,

$$\begin{aligned}\Pr\left(k_1^{(2c-2)} > \left(\frac{7}{8}\right)\frac{(1-\epsilon)^c N}{2}\right) &= \Pr\left(k_1^{(2c-2)} > \left(\frac{7}{6}\right)\left(\frac{3}{4}\right)\frac{(1-\epsilon)^c N}{2}\right) \\ &\leq \left(\frac{e^{1/6}}{(7/6)^{7/6}}\right)^{3(1-\epsilon)^c N/8} \\ &\leq e^{-(1-\epsilon)^c N/2^9}.\end{aligned}$$

So, let us assume now that

$$k_1^{(2c-2)} \leq \left(\frac{7}{8}\right)\frac{(1-\epsilon)^c N}{2},$$

and consider one more random matching. Note that, in this case,

$$E(k_1^{(2c-1)}) \leq \left(\frac{7}{8}\right)\frac{(1-\epsilon)^{c+1}N}{2}.$$

Therefore, by a well-known Chernoff bound,

$$\begin{aligned}\Pr\left(k_1^{(2c-1)} > \frac{(1-\epsilon)^{c+1}N}{2}\right) &= \Pr\left(k_1^{(2c-2)} > \left(\frac{8}{7}\right)\left(\frac{7}{8}\right)\frac{(1-\epsilon)^{c+1}N}{2}\right) \\ &\leq \left(\frac{e^{1/7}}{(8/7)^{8/7}}\right)^{7(1-\epsilon)^{c+1}N/16} \\ &\leq e^{-(1-\epsilon)^{c+1}N/2^{10}}.\end{aligned}$$

Combining all the failure probabilities, as in a union bound, then, establishes the lemma. ∎

By a symmetrical argument, we have similar result for the case of $k \leq (1-\epsilon)N$ zeroes that would wind up in $A_2$ after $c$ random matchings (with compare-exchange operations between the matched pairs). Thus, we have the following.

**Corollary 3.4** *If $A_1$ and $A_2$ are two regions of size $N$ each, then a compare-exchange operation consisting of $2c-1$ random matchings (with compare-exchanges between matched pairs) between $A_1$ and $A_2$ is a $(c+1, 1/2, 0)$-leveraged-splitter with probability at least $1 - (2c-1)e^{-(1-\epsilon)^{c+1}N/2^{10}}$.*

The above lemma and corollary are most useful for cases when the regions are large enough so that the above failure probability is below $O(n^{-\alpha})$, for $\alpha > 1$.

## 3.3 The $(c+1, (2e)^c, 4e\log n)$-Leveraged-Splitter Property

When region sizes or $(1-\epsilon)$ values are too small for Corollary 3.4 to hold, we can use the $(c+1, (2e)^c, 4e\log n)$-leveraged-splitter property of the region-compare operation. As above, we prove this property by assuming, without loss of generality, that we are operating on a zero-one array and by focusing on the $k$ largest elements, that is, the ones. We also note that this particular $(\mu, \alpha, \beta)$-leveraged-splitter property is only useful when $(1-\epsilon) < 1/(2e)$, when considering the $k \leq (1-\epsilon)N$ largest elements (i.e., the ones), so we add this as a condition as well.

**Lemma 3.5 ((Little-Region Lemma))** *Given two regions, $A_1$ and $A_2$, each of size $N$, let $k = k_1 + k_2$, where $k_1$ and $k_2$ are the respective number of ones in $A_1$ and $A_2$. Suppose $k \leq (1-\epsilon)N$, where $\epsilon$ satisfies $(1-\epsilon) < 1/(2e)$. Let $k_1^{(c)}$ be the number of ones in $A_1$ after $c$ region compare-exchange operations. Then*

$$\Pr\left(k_1^{(c)} > \max\{(2e)^c(1-\epsilon)^{c+1}N, 4e\log n\}\right) \leq cn^{-4}.$$

**Proof:** Let us apply an induction argument, based on the number, $c'$, of random matches in a region compare-exchange operation. Consider an inductive claim, which states that after after $c'$ random matchings (with compare-exchange operations),

$$k_1^{(c')} > \max\{(2e)^{c'}(1-\epsilon)^{c'+1}N, \ 4e\log n\},$$

with probability at most $c'n^{-4}$. Thus, with high probability $k_1^{(c')}$ is bounded by the formula on the righthand side. The claim is clearly true by assumption for $c' = 0$. So, suppose the claim is true for $c'$, and let us consider $c'+1$. Since there is at most a $(1-\epsilon)$ fraction of ones in $A_2$,

$$\mu = E(k_1^{(c'+1)}) \leq (2e)^{c'}(1-\epsilon)^{c'+2}N.$$

Moreover, the value $k_1^{(c'+1)}$ can be viewed as the number of ones in a sample without replacement from $A_2$ of size $k_1^{(c')}$. By a theorem of Hoeffding [22], then, the expected value of any convex function of the size of such a sample is bounded by the expected value of that function applied to the size of a similar sample with replacement. Thus, we can apply a Chernoff bound (e.g., see [34, 35]) to this single random matching and pairwise set of compare-exchange operations, to derive

$$\Pr\left(k_1^{(c'+1)} > (1+\delta)\mu\right) < 2^{-\delta\mu},$$

provided $\delta \geq 2e - 1$. Taking $(1+\delta)\mu = 2eM(N)$ implies $\delta \geq 2e - 1$, where $M(N) = (2e)^{c'}(1-\epsilon)^{c'+2}N$; hence, we can bound

$$\Pr\left(k_1^{(c'+1)} > 2eM(N)\right) < 2^{-(2eM(N)-\mu)} \leq 2^{-eM(N)},$$

which also gives us a new bound on $M(N)$ for the next step in the induction. Provided $M(N) \geq 2\log n$, then this (failure) condition holds with probability less than $n^{-4}$. If, on the other hand, $M(N) < 2\log n$, then

$$\Pr\left(k_1^{(c'+1)} > 4e\log n\right) < 2^{-(4e\log n - \mu)} \leq 2^{-2e\log n} < n^{-4}.$$

In this latter case, we can terminate the induction, since repeated applications of the region compare-exchange operation can only improve things. Otherwise, we continue the induction. At some point during the induction, we must either reach $c'+1 = c$, at which point the inductive hypothesis implies the lemma, or we will have $M(N) < 2\log n$, which puts us into the above second case and implies the lemma. ∎

A similar argument applies to the case of the $k$ smallest elements, which gives us the following.

**Corollary 3.6** *If $A_1$ and $A_2$ are two regions of size $N$ each, then a compare-exchange operation consisting of $c$ random matchings (with compare-exchanges between matched pairs) between $A_1$ and $A_2$ is a $(c+1, (2e)^c, 4e\log n)$-leveraged-splitter with probability at least $1 - cn^{-4}$.*

As we noted above, this corollary is only of use for the case when $(1-\epsilon) < 1/(2e)$, where $\epsilon$ is the same parameter as used in the definition of a $(\mu, \alpha, \beta)$-leveraged-splitter.

## 3.4 The $(0, 1/6, 0)$-Leveraged-Splitter Property

The final property we prove is for the $(0, 1/6, 0)$-leveraged-splitter property. As with the other two properties, we consider here the $k$ largest elements, and focus on the case of a zero-one array.

**Lemma 3.7 ((Startup Lemma))** *Given two regions, $A_1$ and $A_2$, each of size $N$, let $k = k_1 + k_2$, where $k_1$ and $k_2$ are the respective number of ones in $A_1$ and $A_2$, and $k \leq N$. Let $k_1^{(c)}$ be the number of ones in $A_1$ after $c$ region compare-exchange operations. Then $k_1^{(4)} \leq N/6$, with very high probability, provided $N$ is $\Omega(\ln n)$.*

**Proof:** The proof involves four consecutive applications of a theorem of Hoeffding [22] that the expected value of any convex function of the size of a sample without replacement is bounded by the expected value of that function applied to the size of a similar sample with replacement. Thus, we can apply a Chernoff bound (e.g., see [34, 35]) to each single random matching and pairwise set of compare-exchange operations, to derive bounds on failure probabilities. As noted above,

$$E\left(k_1^{(1)}\right) = k_1\left(1 - \frac{k_1}{N}\right) \leq \frac{N}{4},$$

since $k_1(1 - k_1/N)$ is maximized at $k_1 = N/2$. Thus, by a well-known Chernoff bound (e.g., see [34, 35]),

$$\Pr\left(k_1^{(1)} > \frac{N}{3}\right) = \Pr\left(k_1^{(1)} > \left(1 + \frac{1}{3}\right)\frac{N}{4}\right) \leq \left(\frac{e^{1/3}}{(4/3)^{4/3}}\right)^{N/4} \leq e^{-N/2^7}.$$

So let us assume $k_1^{(1)} \leq N/3$. Since $k_1(1 - k_1/N)$ is monotonic on $[0, N/2]$,

$$E\left(k_1^{(2)}\right) \leq \frac{N}{3}\left(1 - \frac{1}{3}\right) = \frac{2N}{9}.$$

Thus, by another application of a Chernoff bound,

$$\Pr\left(k_1^{(2)} > \frac{N}{4}\right) = \Pr\left(k_1^{(2)} > \left(1 + \frac{1}{8}\right)\frac{2N}{9}\right) \leq \left(\frac{e^{1/8}}{(9/8)^{9/8}}\right)^{2N/9} \leq e^{-N/2^{11}}.$$

So let us assume $k_1^{(2)} \leq N/4$. Since $k_1(1 - k_1/N)$ is monotonic on $[0, N/2]$,

$$E\left(k_1^{(3)}\right) \leq \frac{N}{4}\left(1 - \frac{1}{4}\right) = \frac{3N}{16}.$$

Thus, again applying a Chernoff bound,

$$\Pr\left(k_1^{(3)} > \frac{N}{5}\right) = \Pr\left(k_1^{(3)} > \left(1 + \frac{1}{15}\right)\frac{3N}{16}\right) \leq \left(\frac{e^{1/15}}{(16/15)^{16/15}}\right)^{3N/16} \leq e^{-N/2^{13}}.$$

So let us assume that $k_1^{(3)} \leq N/5$. Thus, since $k_1(1 - k_1/N)$ is monotonic on $[0, N/2]$,

$$E\left(k_1^{(4)}\right) \leq \frac{N}{5}\left(1 - \frac{1}{5}\right) = \frac{4N}{25}.$$

Thus, by one more application of a Chernoff bound,

$$\Pr\left(k_1^{(4)} > \frac{N}{6}\right) = \Pr\left(k_1^{(4)} > \left(1 + \frac{1}{24}\right)\frac{5N}{25}\right) \leq \left(\frac{e^{1/24}}{(25/24)^{25/24}}\right)^{4N/25} \leq e^{-N/2^{15}}.$$

The proof follows by summing these four failure probabilities and the fact that $N$ is $\Omega(\ln n)$. ■

Of course, the above lemma has an obvious symmetric versions that applies to the number of zeroes on the right side of two regions in a region compare-exchange. Thus, we have the following.

**Corollary 3.8** *If $A_1$ and $A_2$ are two regions of size $N$ each, then a compare-exchange operation consisting of at least 4 random matchings (with compare-exchanges between matched pairs) between $A_1$ and $A_2$ is a $(0, 1/6, 0)$-leveraged-splitter with probability at least $1 - cn^{-4}$, provided $N$ is $\Omega(\ln n)$.*

## 3.5 The $(c+1, 1/2, N/20)$-Leveraged-Splitter Property

Finally, let us consider one additional lemma characterizing region compare-exchange operations. This one is most useful in contexts where $\epsilon$ is relatively small.

**Lemma 3.9 ((Slow-Depletion Lemma))** *Given two regions, $A_1$ and $A_2$, each of size $N$, let $k = k_1 + k_2 \leq (1-\epsilon)N$, where $k_1$ and $k_2$ are the respective number of ones in $A_1$ and $A_2$, and $k \leq N$ and $0 < \epsilon < 1$. Let $k_1^{(c)}$ be the number of ones in $A_1$ after $c$ random matches in a compare-exchange operation. Then*

$$\Pr\left(k_1^{(c)} > \max\{(1-\epsilon)^{c+1}N/2, N/20\}\right) \leq c\, e^{-(1-\epsilon)^{c+1}N/2^{12}}.$$

**Proof:** The proof is by induction on $c$. For the base case, note that if $k_1 = k_1^{(0)} \leq N/20$, then we are done, since random compare-exchanges between $A_1$ and $A_2$ can only improve the number of ones in $A_1$. Furthermore, if $(1-\epsilon)N/2 \leq N/20$, then we are guaranteed to have $k_1^{(1)} \leq N/20$, since the only way one can stay in $A_1$ is if it is matched with a one in $A_2$. So suppose $k_1 > N/20$ and $(1-\epsilon)N/2 > N/20$. In this case, recall that

$$E(k_1^{(1)}) \leq \frac{(1-\epsilon)^2 N}{4}.$$

Thus, combining the theorem of Hoeffding [22] that the expected value of any convex function of the size of a sample without replacement is bounded by the expected value of that function applied to the size of a similar sample with replacement, and a well-known Chernoff bound,

$$\Pr\left(k_1^{(1)} > \frac{(1-\epsilon)^2 N}{2}\right) \leq \left(\frac{e}{4}\right)^{(1-\epsilon)^2 N/4} \leq e^{-(1-\epsilon)^2 N/2^4}.$$

For the inductive step, for $c \geq 2$, let us assume the lemma is true for $c-1$ random matchings and that $(1-\epsilon)^{c-1}N/2 > N/20$. Let us consider the case when previous steps succeeded, in which case, since $x(1-x)$ is monotonic for $x \in [0, 1/2]$,

$$\begin{aligned}
E(k_1^{(c)}) &\leq \frac{(1-\epsilon)^c N}{2}\left((1-\epsilon) - \frac{(1-\epsilon)^c}{2}\right) \\
&= \frac{(1-\epsilon)^{c+1} N}{2}\left(1 - \frac{(1-\epsilon)^{c-1}}{2}\right) \\
&\leq \frac{(1-\epsilon)^{c+1} N}{2}\left(\frac{19}{20}\right).
\end{aligned}$$

Thus, by a well-known Chernoff bound,

$$\begin{aligned}
\Pr\left(k_1^{(c)} > \frac{(1-\epsilon)^{c+1}N}{2}\right) &= \Pr\left(k_1^{(c)} > \left(\frac{20}{19}\right)\left(\frac{19}{20}\right)\frac{(1-\epsilon)^{c+1}N}{2}\right) \\
&\leq \left(\frac{e^{1/19}}{(20/19)^{20/19}}\right)^{19(1-\epsilon)^{c+1}N/40} \\
&\leq e^{-(1-\epsilon)^{c+1}N/2^{12}}.
\end{aligned}$$

If, after $c$ random matchings, $k_1 \leq N/20$ or $(1-\epsilon)^c N/2 \leq N/20$, then we are done, w.v.h.p. Thus, either we satisfy the condition of the lemma or we can continue the induction. Therefore, the proof follows by summing all the failure probabilities. ∎

This implies the following.

**Corollary 3.10** *If $A_1$ and $A_2$ are two regions of size $N$ each, then a compare-exchange operation consisting of $c$ random matchings (with compare-exchanges between matched pairs) between $A_1$ and $A_2$ is a $(c+1, 1/2, N/20)$-leveraged-splitter with probability at least $1 - c\,e^{-(1-\epsilon)^{c+1}N/2^{12}}$.*

# 4 Analyzing the Core Algorithm

Having proven the essential properties of a region compare-exchange operation, consisting of $c$ random matchings (with compare-exchanges between matched pairs), we now turn to the problem of analyzing the core part of our randomized Shellsort algorithm.

## 4.1 A Probabilistic Zero-One Principle

We begin our analysis with a ***probabilistic version of the zero-one principle***.

**Lemma 4.1** *If a randomized data-oblivious sorting algorithm sorts any binary array of size $n$ with failure probability at most $\epsilon$, then it sorts any arbitrary array of size $n$ with failure probability at most $\epsilon(n+1)$.*

**Proof:** The lemma[2] follows from the proof of Theorem 3.3 by Rajasekaran and Sen [39], which itself is based on the justification of Knuth [26] for the deterministic version of the zero-one principle for sorting networks. The essential fact is that an arbitrary $n$-element input array, $A$, has, via monotonic bijections, at most $n+1$ corresponding $n$-length binary arrays, such that $A$ is sorted correctly by a data-oblivious algorithm, $\mathcal{A}$, if and only if every bijective binary array is sorted correctly by $\mathcal{A}$. (See Rajasekaran and Sen [39] or Knuth [26] for the proof of this fact.) ∎

Note that this lemma is only of practical use for randomized data-oblivious algorithms that have failure probabilities of at most $O(n^{-a})$, for some constant $a > 1$. We refer to such algorithms as succeeding **with very high probability**. Fortunately, our analysis shows that our randomized Shellsort algorithm will $\alpha \operatorname{polylog}(n)$-near-sort a binary array with very high probability.

## 4.2 Bounding Dirtiness after each Iteration

In the $d$-th iteration of our core algorithm, we partition the array $A$ into $2^d$ regions, $A_0, A_1, \ldots, A_{2^d-1}$, each of size $n/2^d$. Moreover, each iteration splits a region from the previous iteration into two equal-sized halves. Thus, the algorithm can be visualized in terms of a complete binary tree, $B$, with $n$ leaves. The root of $B$ corresponds to a region consisting of the entire array $A$ and each leaf of $B$ corresponds to an individual cell, $a_i$, in $A$, of size 1. Each internal node $v$ of $B$ at depth $d$ corresponds with a region, $A_i$, created in the $d$-th iteration of the algorithm, and the children of $v$ are associated with the two regions that $A_i$ is split into during the $(d+1)$-st iteration. (See Figure 3.)

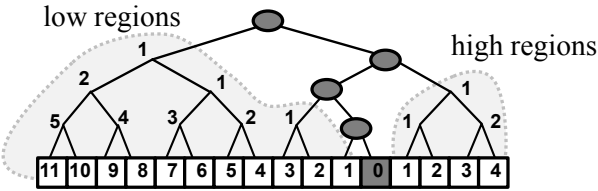

Figure 3: The binary tree, $B$, and the distance of each region from the mixed region (shown in dark gray).

The desired output, of course, is to have each leaf value, $a_i = 0$, for $i < n - k$, and $a_i = 1$, otherwise. We therefore refer to the transition from cell $n - k - 1$ to cell $n - k$ on the last level of $B$ as the **crossover** point.

---
[2]A similar lemma is provided by Blackston and Ranade [7], but they omit the proof.

We refer to any leaf-level region to the left of the crossover point as a **low** region and any leaf-level region to the right of the crossover point as a **high** region. We say that a region, $A_i$, corresponding to an internal node $v$ of $B$, is a **low** region if all of $v$'s descendents are associated with low regions. Likewise, a region, $A_i$, corresponding to an internal node $v$ of $B$, is a **high** region if all of $v$'s descendents are associated with high regions. Thus, we desire that low regions eventually consist of only zeroes and high regions eventually consist of only ones. A region that is neither high nor low is **mixed**, since it is an ancestor of both low and high regions. Note that there are no mixed leaf-level regions, however.

Also note that, since our randomized Shellsort algorithm is data-oblivious, the algorithm doesn't take any different behavior depending on whether is a region is high, low, or mixed. Nevertheless, since the region-compare operation is w.v.h.p. a $(\mu, \alpha, \beta)$-leveraged-splitter, for each of the $(\mu, \alpha, \beta)$ tuples, $(c+1, 1/2, 0)$, $(c+1, (2e)^c, 4e\log n)$, and $(0, 1/6, 0)$, we can reason about the actions of our algorithm on different regions in terms of any one of these tuples.

With each high (resp., low) region, $A_i$, define the **dirtiness** of $A_i$ to be the number of zeroes (resp., ones) that are present in $A_i$, that is, values of the wrong type for $A_i$. With each region, $A_i$, we associate a dirtiness bound, $\delta(A_i)$, which is a desired upper bound on the dirtiness of $A_i$.

For each region, $A_i$, at depth $d$ in $B$, let $j$ be the number of regions between $A_i$ and the crossover point or mixed region on that level. That is, if $A_i$ is a low leaf-level region, then $j = n - k - i - 1$, and if $A_i$ is a high leaf-level region, then $j = j - n + k$. We define the **desired dirtiness bound**, $\delta(A_i)$, of $A_i$ as follows:

- If $j \geq 2$, then
$$\delta(A_i) = \frac{n}{2^{d+j+3}}.$$

- If $j = 1$, then
$$\delta(A_i) = \frac{n}{5 \cdot 2^d}.$$

- If $A_i$ is a mixed region, then
$$\delta(A_i) = |A_i|.$$

Thus, every mixed region trivially satisfies its desired dirtiness bound.

Because of our need for a high probability bound, we will guarantee that each region $A_i$ satisfies its desired dirtiness bound, w.v.h.p., only if $\delta(A_i) \geq 12e \log n$. If $\delta(A_i) < 12e \log n$, then we say $A_i$ is an **extreme** region, for, during our algorithm, this condition implies that $A_i$ is relatively far from the crossover point. (Please see Figure 4, for an illustration of the "zones of order" that are defined by the low, high, mixed, and extreme regions in $A$.)

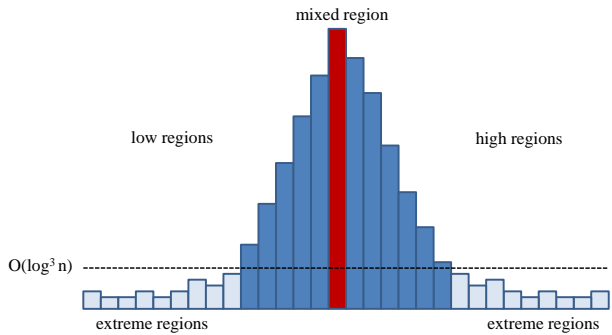

Figure 4: An example histogram of the dirtiness of the different kinds of regions, as categorized by the analysis of the randomized Shellsort algorithm. By the inductive claim, the distribution of dirtiness has exponential tails with polylogarithmic ends.

We will show that the total dirtiness of all extreme regions is $O(\log^3 n)$ w.v.h.p. Thus, we can terminate our analysis when the number and size of the non-extreme regions is polylog($n$), at which point the array $A$ will be $O(\text{polylog}(n))$-near-sorted w.v.h.p. Throughout this analysis, we make repeated use of the following simple but useful lemma.

**Lemma 4.2** *Suppose $A_i$ is a low (resp., high) region and $\Delta$ is the cumulative dirtiness of all regions to the left (resp., right) of $A_i$. Then any region compare-exchange pass over $A$ can increase the dirtiness of $A_i$ by at most $\Delta$.*

**Proof:** If $A_i$ is a low (resp., high) region, then its dirtiness is measured by the number of ones (resp., zeroes) it contains. During any region compare-exchange pass, ones can only move right, exchanging themselves with zeroes, and zeroes can only move left, exchanging themselves with ones. Thus, the only ones that can move into a low region are those to the left of it and the only zeroes that can move into a high region are those to the right of it. ∎

### 4.3 An Inductive Argument

The inductive claim we wish to show holds with very high probability is the following.

**Claim 4.3** *After iteration $d$, for each region $A_i$, the dirtiness of $A_i$ is at most $\delta(A_i)$, provided $A_i$ is not extreme. The total dirtiness of all extreme regions is at most $12ed \log^2 n$.*

Let us begin at the point when the algorithm creates the first two regions, $A_1$ and $A_2$. Suppose that $k \le n - k$, where $k$ is the number of ones, so that $A_1$ is a low region and $A_2$ is either a high region (i.e., if $k = n-k$) or $A_2$ is mixed (the case when $k > n-k$ is symmetric). Let $k_1$ (resp., $k_2$) denote the number of ones in $A_1$ (resp., $A_2$), so $k = k_1 + k_2$. By the Startup Lemma (3.7), the dirtiness of $A_1$ will be at most $n/12$, with very high probability, since the region compare-exchange operation is a $(0, 1/6, 0)$-leveraged-splitter. Note that this satisfies the desired dirtiness of $A_1$, since $\delta(A_1) = n/10$ in this case. A similar argument applies to $A_2$ if it is a high region, and if $A_2$ is mixed, it trivially satisfies its desired dirtiness bound. Also, assuming $n$ is large enough, there are no extreme regions (if $n$ is so small that $A_1$ is extreme, we can immediately switch to the postprocessing cleanup phase). Thus, we satisfy the base case of our inductive argument—the dirtiness bounds for the two children of the root of $B$ are satisfied with (very) high probability, and similar arguments prove the inductive claim for iterations 2 and 3.

Let us now consider a general inductive step. Let us assume that, with very high probability, we have satisfied Claim 4.3 for the regions on level $d \ge 3$ and let us now consider the transition to level $d+1$. In addition, we terminate this line of reasoning when the region size, $n/2^d$, becomes less than $16e^2 \log^6 n$, at which point $A$ will be $O(\text{polylog}(n))$-near-sorted, with very high probability, by Claim 4.3 and Lemma 4.1.

#### 4.3.1 Extreme Regions

Let us begin with the bound for the dirtiness of extreme regions in iteration $d+1$. Note that, by Lemma 4.2, regions that were extreme after iteration $d$ will be split into regions in iteration $d+1$ that contribute no new amounts of dirtiness to pre-existing extreme regions. That is, extreme regions get split into extreme regions. Thus, the new dirtiness for extreme regions can come only from regions that were not extreme after iteration $d$ that are now splitting into extreme regions in iteration $d+1$, which we call ***freshly extreme*** regions. Suppose, then, that $A_i$ is such a region, say, with a parent, $A_p$, which is $j$ regions from the mixed region on level $d$. Then the desired dirtiness bound of $A_i$'s parent region, $A_p$, is $\delta(A_p) = n/2^{d+j+3} \ge 12e \log n$, by Claim 4.3, since $A_p$ is not extreme. $A_p$ has (low-region) children, $A_i$ and $A_{i+1}$, that have desired dirtiness bounds of $\delta(A_i) = n/2^{d+1+2j+4}$ or $\delta(A_i) = n/2^{d+1+2j+3}$ and of $\delta(A_{i+1}) = n/2^{d+1+2j+3}$ or $\delta(A_{i+1}) = n/2^{d+1+2j+2}$, depending on whether the mixed region on level $d+1$ has an odd or even index. Moreover, $A_i$ (and possibly $A_{i+1}$) is freshly extreme, so $n/2^{d+1+2j+4} < 12e \log n$, which implies that $j > (\log n - d - \log \log n - 10)/2$. Nevertheless, note also that there are $O(\log n)$ new regions on this level that are just now becoming extreme, since $n/2^d > 16e^2 \log^6 n$ and $n/2^{d+j+3} \ge 12e \log n$ implies $j \le \log n - d$. So let us consider the two new regions, $A_i$ and $A_{i+1}$, in turn, and how the shaker pass effects them (for after that they will collectively satisfy the extreme-region part of Claim 4.3).

- **Region $A_i$:** Consider the worst case for $\delta(A_i)$, namely, that $\delta(A_i) = n/2^{d+1+2j+4}$. Since $A_i$ is a left child of $A_p$, $A_i$ could get at most $n/2^{d+j+3} + 12ed\log^2 n$ ones from regions left of $A_i$, by Lemma 4.2. In addition, $A_i$ and $A_{i+1}$ could inherit at most $\delta(A_p) = n/2^{d+j+3}$ ones from $A_p$. Thus, if we let $N$ denote the size of $A_i$, i.e., $N = n/2^{d+1}$, then $A_i$ and $A_{i+1}$ together have at most $N/2^{j+1} + 3N^{1/2} \leq N/2^j$ ones, since we stop the induction when $N < 16e^2 \log^6 n$. By the Little-Region Lemma (3.5), the following condition holds with probability at least $1 - cn^{-4}$,

$$k_1^{(c)} \leq \max\{(2e)^c(1-\epsilon)^{c+1}N,\, 4e\log n\},$$

where $k_1^{(c)}$ is the number of one left in $A_i$ after $c$ region compare-exchanges with $A_{i+1}$, since the region compare-exchange operation is a $(c+1, (2e)^c, 4e\log n)$-leveraged-splitter. Note that, if $k_1^{(c)} \leq 4e\log n$, then we have satisfied the desired dirtiness for $A_i$. Alternatively, so long as $c \geq 4$, and $j \geq 5$, then w.v.h.p.,

$$\begin{aligned} k_1^{(c)} &\leq (2e)^c(1-\epsilon)^{c+1}N \leq \frac{(2e)^c n}{2^{d+1+j(c+1)}} \\ &\leq \frac{n}{2^{d+1+2j+3}} < 12e\log n = \delta(A_i). \end{aligned}$$

- **Region $A_{i+1}$:** Consider the worst case for $\delta(A_{i+1})$, namely $\delta(A_{i+1}) = n/2^{d+1+2j+3}$. Since $A_{i+1}$ is a right child of $A_p$, $A_{i+1}$ could get at most $n/2^{d+j+3} + 12ed\log^2 n$ ones from regions left of $A_{i+1}$, by Lemma 4.2, plus $A_{i+1}$ could inherit at most $\delta(A_p) = n/2^{d+j+3}$ ones from $A_p$ itself. In addition, since $j > 2$, $A_{i+2}$ could inherit at most $n/2^{d+j+2}$ ones from its parent. Thus, if we let $N$ denote the size of $A_{i+1}$, i.e., $N = n/2^{d+1}$, then $A_{i+1}$ and $A_{i+2}$ together have at most $N/2^j + 3N^{1/2} \leq N/2^{j-1}$ ones, since we stop the induction when $N < 16e^2 \log^6 n$. By the Little-Region Lemma (3.5), the following condition holds with probability at least $1 - cn^{-4}$,

$$k_1^{(c)} \leq \max\{(2e)^c(1-\epsilon)^{c+1}N,\, 4e\log n\},$$

where $k_1^{(c)}$ is the number of ones left in $A_{i+1}$ after $c$ region compare-exchange operations, since the region compare-exchange operation is a $(c+1, (2e)^c, 4e\log n)$-leveraged-splitter. Note that, if $k_1^{(c)} \leq 4e\log n$, then we have satisfied the desired dirtiness bound for $A_{i+1}$. Alternatively, so long as $c \geq 4$, and $j \geq 6$,

$$\begin{aligned} k_1^{(c)} &\leq (2e)^c(1-\epsilon)^{c+1}N \leq \frac{(2e)^c n}{2^{d+1+(j-1)(c+1)}} \\ &\leq \frac{n}{2^{d+1+2j+2}} < 12e\log n = \delta(A_{i+1}). \end{aligned}$$

Therefore, if a low region $A_i$ or $A_{i+1}$ becomes freshly extreme in iteration $d+1$, then, w.v.h.p., its dirtiness is at most $12e\log n$. Since there are at most $\log n$ freshly extreme regions created in iteration $d+1$, this implies that the total dirtiness of all extreme low regions in iteration $d+1$ is at most $12e(d+1)\log^2 n$, w.v.h.p., after the right-moving shaker pass, by Claim 4.3. Likewise, by symmetry, a similar claim applies to the high regions after the left-moving shaker pass. Moreover, by Lemma 4.2, these extreme regions will continue to satisfy Claim 4.3 after this.

### 4.3.2 Non-extreme Regions not too Close to the Crossover Point

Let us now consider non-extreme regions on level $d+1$ that are at least two regions away from the crossover point on level $d+1$. Consider, wlog, a low region, $A_p$, on level $d$, which is $j$ regions from the crossover point on level $d$, with $A_p$ having (low-region) children, $A_i$ and $A_{i+1}$, that have desired dirtiness bounds of $\delta(A_i) = n/2^{d+1+2j+4}$ or $\delta(A_i) = n/2^{d+1+2j+3}$ and of $\delta(A_{i+1}) = n/2^{d+1+2j+3}$ or $\delta(A_{i+1}) = n/2^{d+1+2j+2}$, depending on whether the mixed region on level $d+1$ has an odd or even index. By Lemma 4.2, if we can show w.v.h.p. that the dirtiness of each such $A_i$ (resp., $A_{i+1}$) is at most $\delta(A_i)/3$ (resp., $\delta(A_{i+1})/3$), after the shaker pass, then no matter how many more ones come into $A_i$ or $A_{i+1}$ from the left during the rest of iteration $d+1$, they will satisfy their desired dirtiness bounds.

Let us consider the different region types (always taking the most difficult choice for each desired dirtiness in order to avoid additional cases):

- **Type 1:** $\delta(A_i) = n/2^{d+1+2j+4}$, with $j \geq 2$. Since $A_i$ is a left child of $A_p$, $A_i$ could get at most $n/2^{d+j+3} + 12ed \log^2 n$ ones from regions left of $A_i$, by Lemma 4.2. In addition, $A_i$ and $A_{i+1}$ could inherit at most $\delta(A_p) = n/2^{d+j+3}$ ones from $A_p$. Thus, if we let $N$ denote the size of $A_i$, i.e., $N = n/2^{d+1}$, then $A_i$ and $A_{i+1}$ together have at most $N/2^{j+1} + 3N^{1/2} \leq N/2^j$ ones, since we stop the induction when $N < 16e^2 \log^6 n$. If $(1-\epsilon)^{c+1} N/2^{10} \ln(2c-1) \geq 4 \ln n$, then, by the Fast-Depletion Lemma (3.3), the following condition holds with probability at least $1 - n^{-4}$, provided $c \geq 4$:

$$k_1^{(2c-1)} \leq \frac{(1-\epsilon)^{c+1} N}{2} \leq \frac{n}{2^{d+1+j(c+1)+1}}$$
$$\leq \frac{n}{3 \cdot 2^{d+1+2j+4}} = \delta(A_i)/3,$$

where $k_1^{(c)}$ is the number of ones left in $A_i$ after $c$ region compare-exchange operations, since the region compare-exchange operation is a $(c+1, 1/2, 0)$-leveraged-splitter. If, on the other hand, $(1-\epsilon)^{c+1} N/2^{10} \ln(2c-1) < 4 \ln n$, then $j$ is $\Omega(\log \log n)$, so we can assume $j \geq 6$, and, by the Little-Region Lemma (3.5), the following condition holds with probability at least $1 - cn^{-4}$ in this case:

$$k_1^{(c)} \leq \max\{(2e)^c (1-\epsilon)^{c+1} N, \, 4e \log n\},$$

since the region compare-exchange operation is a $(c+1, (2e)^c, 4e \log n)$-leveraged-splitter. Note that, since $A_i$ is not extreme, if $k_1^{(c)} \leq 4e \log n$, then $k_1^{(c)} \leq \delta(A_i)/3$. Alternatively, so long as $c \geq 4$, then, w.v.h.p.,

$$k_1^{(c)} \leq (2e)^c (1-\epsilon)^{c+1} N \leq \frac{(2e)^c n}{2^{d+1+j(c+1)}}$$
$$\leq \frac{n}{3 \cdot 2^{d+1+2j+4}} = \delta(A_i)/3.$$

- **Type 2:** $\delta(A_{i+1}) = n/2^{d+1+2j+3}$, with $j > 2$. Since $A_{i+1}$ is a right child of $A_p$, $A_{i+1}$ could get at most $n/2^{d+j+3} + 12ed \log^2 n$ ones from regions left of $A_{i+1}$, by Lemma 4.2, plus $A_{i+1}$ could inherit at most $\delta(A_p) = n/2^{d+j+3}$ ones from $A_p$. In addition, since $j > 2$, $A_{i+2}$ could inherit at most $n/2^{d+j+2}$ ones from its parent. Thus, if we let $N$ denote the size of $A_{i+1}$, i.e., $N = n/2^{d+1}$, then $A_{i+1}$ and $A_{i+2}$ together have at most $N/2^j + 3N^{1/2} \leq N/2^{j-1}$ ones, since we stop the induction when $N < 16e^2 \log^6 n$. If $(1-\epsilon)^{c+1} N/2^{10} \ln(2c-1) \geq 4 \ln n$, then, by the Fast-Depletion Lemma (3.3), the following condition holds with probability at least $1 - n^{-4}$, for a suitably-chosen constant $c$,

$$k_1^{(2c-1)} \leq \frac{(1-\epsilon)^{c+1} N}{2} \leq \frac{n}{2^{d+1+(j-1)(c+1)+1}}$$
$$\leq \frac{n}{3 \cdot 2^{d+1+2j+3}} = \delta(A_{i+1})/3,$$

where $k_1^{(c)}$ is the number of ones left in $A_{i+1}$ after $c$ region compare-exchange operations. If, on the other hand, $(1-\epsilon)^{c+1} N/2^{10} \ln(2c-1) < 4 \ln n$, then $j$ is $\Omega(\log \log n)$, so we can now assume $j \geq 6$, and, by the Little-Region Lemma (3.5), the following condition holds with probability at least $1 - cn^{-4}$:

$$k_1^{(c)} \leq \max\{(2e)^c (1-\epsilon)^{c+1} N, \, 4e \log n\}.$$

Note that, since $A_i$ is not extreme, if $k_1^{(c)} \leq 4e \log n$, then $k_1^{(c)} \leq \delta(A_{i+1})/3$. Thus, we can choose constant $c$ so that

$$k_1^{(c)} \leq (2e)^c (1-\epsilon)^{c+1} N \leq \frac{(2e)^c n}{2^{d+1+(j-1)(c+1)}}$$
$$\leq \frac{n}{3 \cdot 2^{d+1+2j+3}} = \delta(A_{i+1})/3.$$

- **Type 3:** $\delta(A_{i+1}) = n/2^{d+1+2j+3}$, with $j = 2$. Since $A_{i+1}$ is a right child of $A_p$, $A_{i+1}$ could get at most $n/2^{d+j+3} + 12ed \log^2 n$ ones from regions left of $A_{i+1}$, by Lemma 4.2, plus $A_{i+1}$ could inherit at most $\delta(A_p) = n/2^{d+j+3}$ ones from $A_p$. In addition, since $j = 2$, $A_{i+2}$ could inherit at most $n/(5 \cdot 2^d)$ ones from its parent. Thus, if we let $N$ denote the size of $A_{i+1}$, i.e., $N = n/2^{d+1}$, then $A_{i+1}$ and $A_{i+2}$ together have at most $N/2^{j+1} + 2N/5 + 3N^{1/2} \leq 3N/5$ ones, since we stop the induction when $N < 16e^2 \log^6 n$. In addition, note that this also implies that as long as $c$ is a constant, $(1-\epsilon)^{c+1} N/2^{10} \ln(2c-1) \geq 4 \ln n$. Thus, by the Fast-Depletion Lemma (3.3), we can choose constant $c$ so that the following condition holds with probability at least $1 - n^{-4}$:

$$\begin{aligned} k_1^{(2c-1)} &\leq \frac{(1-\epsilon)^{c+1} N}{2} \leq \frac{3^{c+1} n}{5^{c+1} 2^{d+2}} \\ &\leq \frac{n}{3 \cdot 2^{d+1+2j+3}} = \delta(A_{i+1})/3, \end{aligned}$$

where $k_1^{(c)}$ is the number of ones left in $A_{i+1}$ after $c$ region compare-exchange operations.

- **Type 4:** $\delta(A_i) = n/2^{d+1+2j+4}$, with $j = 1$. Since $A_i$ is a left child of $A_p$, $A_i$ could get at most $n/2^{d+j+2} + 12ed \log^2 n$ ones from regions left of $A_i$, by Lemma 4.2, plus $A_i$ and $A_{i+1}$ could inherit at most $\delta(A_p) = n/(5 \cdot 2^d)$ ones from $A_p$. Thus, if we let $N$ denote the size of $A_i$, i.e., $N = n/2^{d+1}$, then $A_i$ and $A_{i+1}$ together have at most $N/2^{j+1} + 2N/5 + 3N^{1/2} \leq 7N/10$ ones, since we stop the induction when $N < 16e^2 \log^6 n$. In addition, note that this also implies that as long as $c$ is a constant, $(1-\epsilon)^{c+1} N/2^{10} \ln(2c-1) \geq 4 \ln n$. Thus, by the Fast-Depletion Lemma (3.3), the following condition holds with probability at least $1 - n^{-4}$, for a suitably-chosen constant $c$,

$$\begin{aligned} k_1^{(2c-1)} &\leq \frac{(1-\epsilon)^{c+1} N}{2} \leq \frac{7^{c+1} n}{10^{c+1} 2^{d+2}} \\ &\leq \frac{n}{3 \cdot 2^{d+1+2j+4}} = \delta(A_i)/3, \end{aligned}$$

where $k_1^{(c)}$ is the number of ones left in $A_i$ after $c$ region compare-exchange operations.

Thus, $A_i$ and $A_{i+1}$ satisfy their respective desired dirtiness bounds w.v.h.p., provided they are at least two regions from the mixed region or crossover point.

### 4.3.3 Regions near the Crossover Point

Consider now regions near the crossover point. That is, each region with a parent that is mixed, bordering the crossover point, or next to a region that either contains or borders the crossover point. Let us focus specifically on the case when there is a mixed region on levels $d$ and $d+1$, as it is the most difficult of these scenarios.

So, having dealt with all the other regions, which have their desired dirtiness satisfied after the shaker pass, we are left with four regions near the crossover point, which we will refer to as $A_1$, $A_2$, $A_3$, and $A_4$. One of $A_2$ or $A_3$ is mixed—without loss of generality, let us assume $A_3$ is mixed. At this point in the algorithm, we perform a brick-type pass, which, from the perspective of these four regions, amounts to a complete 4-tournament. Note that, by the results of the shaker pass (which were proved above), we have at this point pushed to these four regions all but at most $n/2^{d+7} + 12e(d+1) \log^2 n$ of the ones and all but at most $n/2^{d+6} + 12e(d+1) \log^2 n$ of the zeroes. Moreover, these bounds will continue to hold (and could even improve) as we perform the different steps of the brick-type pass. Thus, at the beginning of the 4-tournament for these four regions, we know that the four regions hold between $2N - N/32 - 3N^{1/2}$ and $3N + N/64 + 3N^{1/2}$ zeroes and between $N - N/64 - 3N^{1/2}$ and $2N + N/32 + 3N^{1/2}$ ones, where $N = n/2^{d+1} > 16e^2 \log^6 n$. For each region compare-exchange operation, we distinguish three possible outcomes:

- **balanced**: $A_i$ and $A_{i+j}$ have between $31N/32$ and $33/32$ zeroes (and ones). In this case, the Startup Lemma (3.7) implies that $A_i$ will get at least $31N/32 - N/6$ zeroes and at most $N/32 + N/6$ ones, and $A_{i+j}$ will get at least $31N/32 - N/6$ ones and at most $N/32 + N/6$ zeroes, w.v.h.p.

- **0-heavy**: $A_i$ and $A_{i+j}$ have at least $33N/32$ zeroes. In this case[3], by the Slow-Depletion Lemma (3.9), $A_i$ will get at most $N/20$ ones, w.v.h.p., with appropriate choice for $c$.

---
[3] The constant factor can be improved somewhat by first applying the Startup Lemma and then applying the Slow-Depletion Lemma.

- **1-*heavy*:** $A_i$ and $A_{i+j}$ have at least $33N/32$ ones. In this case, by the Slow-Depletion Lemma (3.9), $A_{i+j}$ will get at most $N/20$ zeroes, w.v.h.p., with appropriate choice for $c$.

Let us focus on the four regions, $A_1$, $A_2$, $A_3$, and $A_4$, and consider the region compare-exchange operations that each region participates in as a part of the 4-tournament for these four.

- $A_1$: this region is compared to $A_4$, $A_3$, and $A_2$, in this order. If the first of these is 0-heavy, then we already will satisfy $A_1$'s desired dirtiness bound (which can only improve after this). If the first of these comparisons is balanced, on the other hand, then $A_1$ ends up with at least $31N/32 - N/6 \approx 0.802N$ zeroes (and $A_4$ will have at most $N/32 + N/6 \approx 0.198N$). Since there are at least $2N - N/32 - 3N^{1/2} \approx 1.9N$ zeroes distributed among the four regions, this forces one of the comparisons with $A_3$ or $A_2$ to be 0-heavy, which will cause $A_1$ to satisfy its desired dirtiness.

- $A_2$: this region is compared to $A_4$, $A_1$, and $A_3$, in this order. Note, therefore, that it does its comparisons with $A_4$ and $A_3$ after $A_1$. But even if $A_1$ receives $N$ zeroes, there are still at least $31N/32 - 3N^{1/2}$ zeroes that would be left. Thus, even under this worst-case scenario (from $A_2$'s perspective), the comparisons with $A_2$ and $A_4$ will be either balanced or 1-heavy. If one of them is balanced (and even if $A_1$ is full of zeroes), then $A_2$ gets at least $31N/32 - N/6 \approx 0.802N$ zeroes. If they are both 1-heavy, then $A_2$ and $A_3$ end up with at most $N/20$ zeroes each, which leaves $A_2$ with at least $31N/32 - N/10 \approx 0.869N$ zeroes, w.v.h.p.

- $A_3$: by assumption, $A_3$ is mixed, so it automatically satisfies its desired dirtiness bound.

- $A_4$: this region is compared to $A_1$, $A_2$, and $A_3$, in this order. If any of these is balanced or 1-heavy, then we satisfy the desired dirtiness bound for $A_4$. If they are all 0-heavy, then each of them ends up with at most $N/20$ ones each, which implies that $A_4$ ends up with at least $N - N/64 - 3N/20 - 3N^{1/2} \approx 0.81N$ ones, w.v.h.p., which also satisfies the desired dirtiness bound for $A_4$.

Thus, after the brick-type pass of iteration $d+1$, we will have satisfied Claim 4.3 w.v.h.p. In particular, we have proved that each region satisfies Claim 4.3 after iteration $d+1$ with a failure probability of at most $O(n^{-4})$, for each region compare-exchange operation we perform. Thus, since there are $O(n)$ such regions per iteration, this implies any iteration will fail with probability at most $O(n^{-3})$. Therefore, since there are $O(\log n)$ iterations, and we lose only an $O(n)$ factor in our failure probability when we apply the probabilistic zero-one principle (Lemma 4.1), when we complete the first phase of our randomized Shellsort algorithm, the array $A$ will be $O(\text{polylog}(n))$-near-sorted w.v.h.p., in which case the postprocessing step will complete the sorting of $A$.

## 5 Implementation and Experiments

As an existence proof for its ease of implementation, we provide a complete Java program for randomized Shellsort in Figure 5.

Given this implementation, we explored empirically the degree to which the success of the algorithm depends on the constant $c$, which indicates the number of times to perform random matchings in a region compare-exchange operation. We began with $c = 1$, with the intention of progressively increasing $c$ until we determined the value of $c$ that would lead to failure rate of at most $0.1\%$ in practice. Interestingly, however, $c = 1$ already achieved over a $99.9\%$ success rate in all our experiments.

So, rather than incrementing $c$, we instead kept $c = 1$ and tested the degree to which the different parts of the brick-type pass were necessary, since previous experimental work exists for shaker passes [9, 24, 23, 48]. The first experiment tested the failure percentages of 10,000 runs of randomized Shellsort on random inputs of various sizes, while optionally omitting the various parts of the brick pass while keeping $c = 1$ for region compare-exchange operations and always doing the shaker pass. The failure rates were as follows:

```java
import java.util.*;
public class ShellSort {
  public static final int C=4; // number of region compare-exchange repetitions
  public static void exchange(int[] a, int i, int j) {
    int temp = a[i];
    a[i] = a[j];
    a[j] = temp;
  }
  public static void compareExchange(int[] a, int i, int j) {
    if (((i < j) && (a[i] > a[j])) || ((i > j) && (a[i] < a[j])))                    10
      exchange(a, i, j);
  }
  public static void permuteRandom(int a[], MyRandom rand) {
    for (int i=0; i<a.length; i++) // Use the Knuth random perm. algorithm
      exchange(a, i, rand.nextInt(a.length−i)+i);
  }
  // compare-exchange two regions of length offset each
  public static void compareRegions(int[] a, int s, int t, int offset, MyRandom rand) {
    int mate[] = new int[offset]; // index offset array
    for (int count=0; count<C; count++) { // do C region compare-exchanges           20
      for (int i=0; i<offset; i++) mate[i] = i;
      permuteRandom(mate,rand); // comment this out to get a deterministic Shellsort
      for (int i=0; i<offset; i++)
        compareExchange(a, s+i, t+mate[i]);
    }
  }
  public static void randomizedShellSort(int[] a) {
    int n = a.length; // we assume that n is a power of 2
    MyRandom rand = new MyRandom(); // random number generator (not shown)
    for (int offset = n/2; offset > 0; offset /= 2) {                                30
      for (int i=0; i < n − offset; i += offset) // compare-exchange up
        compareRegions(a,i,i+offset,offset,rand);
      for (int i=n−offset; i >= offset; i −= offset) // compare-exchange down
        compareRegions(a,i−offset,i,offset,rand);
      for (int i=0; i < n−3*offset; i += offset) // compare 3 hops up
        compareRegions(a,i,i+3*offset,offset,rand);
      for (int i=0; i < n−2*offset; i += offset) // compare 2 hops up
        compareRegions(a,i,i+2*offset,offset,rand);
      for (int i=0; i < n; i += 2*offset) // compare odd-even regions
        compareRegions(a,i,i+offset,offset,rand);                                    40
      for (int i=offset; i < n−offset; i += 2*offset) // compare even-odd regions
        compareRegions(a,i,i+offset,offset,rand);
    }
  }
}
```

Figure 5: Our randomized Shellsort algorithm in Java. Note that, just by commenting out the call to permuteRandom, on line 22, in compareRegions, this becomes a deterministic Shellsort implementation.

| $n$ | no brick pass | no short jumps | no long jumps | full pass |
|---|---|---|---|---|
| 128 | 68.18% | 33.92% | 0.01% | 0% |
| 256 | 93.27% | 60.11% | 0% | 0% |
| 512 | 99.86% | 85.62% | 0% | 0% |
| 1024 | 100.00% | 98.27% | 0.01% | 0% |
| 2048 | 100.00% | 99.98% | 0.03% | 0% |
| 4096 | 100.00% | 100.00% | 0.17% | 0% |
| 8192 | 100.00% | 100.00% | 0.14% | 0% |
| 16384 | 100.00% | 100.00% | 0.35% | 0% |
| 32768 | 100.00% | 100.00% | 0.71% | 0% |
| 65536 | 100.00% | 100.00% | 1.53% | 0% |
| 131072 | 100.00% | 100.00% | 2.55% | 0% |
| 262144 | 100.00% | 100.00% | 5.29% | 0% |
| 524288 | 100.00% | 100.00% | 10.88% | 0.01% |
| 1048576 | 100.00% | 100.00% | 21.91% | 0% |

Thus, the need for brick-type passes when $c = 1$ is established empirically from this experiment, with a particular need for the short jumps (i.e., the ones between adjacent regions), but with long jumps still being important.

We next continued the experiment on larger arrays, testing 1,000 runs of randomized Shellsort on random inputs of various sizes, tabulating the failure percentages for performing short-jumps only and full-brick passes. The failure rates were as follows:

| $n$ | short-jumps only | full brick pass |
|---|---|---|
| 2097152 | 34.9% | 0% |
| 4194304 | 59.4% | 0% |

# 6 Conclusion and Open Problems

We have given a simple, randomized Shellsort algorithm that runs in $O(n \log n)$ time and sorts any given input permutation with very high probability. This algorithm can alternatively be viewed as a randomized construction of a simple compare-exchange network that has $O(n \log n)$ size and sorts with very high probability. Its depth is not as asymptotically shallow as the AKS sorting network [1] and its improvements [36, 45], but its constant factors are much smaller and it is quite simple, making it an alternative to the randomized sorting-network construction of Leighton and Plaxton [29]. Some open questions and directions for future work include the following:

- For what values of $\mu$, $\alpha$, and $\beta$ can one deterministically and effectively construct $(\mu, \alpha, \beta)$-leveraged-splitters?

- Is there a simple deterministic $O(n \log n)$-sized sorting network?

- Can the randomness needed for a randomized Shellsort algorithm be reduced to a polylogarithmic number of bits while retaining a very high probability of sorting?

- Can the shaker pass in our randomized Shellsort algorithm be replaced by a lower-depth network, thereby achieving polylogarithmic depth while keeping the overall $O(n \log n)$ size and very high probability of sorting?

- Can the constant factors in the running time for a randomized Shellsort algorithm be reduced to be at most 2 while still maintaining the overall $O(n \log n)$ size and very high probability of sorting?

## Acknowledgments

This research was supported in part by the National Science Foundation under grants 0724806, 0713046, and 0847968, and by the Office of Naval Research under MURI grant N00014-08-1-1015. We are thankful to Bob Sedgewick and an anonymous referee for several of the open problems.

# References

[1] M. Ajtai, J. Komlós, and E. Szemerédi. Sorting in $c \log n$ parallel steps. *Combinatorica*, 3:1–19, 1983.

[2] S. Arora, T. Leighton, and B. Maggs. On-line algorithms for path selection in a nonblocking network. In *STOC '90: Proceedings of the 22nd ACM Symposium on Theory of computing*, pages 149–158, New York, NY, USA, 1990. ACM.

[3] S. Assaf and E. Upfal. Fault tolerant sorting networks. *SIAM J. Discrete Math.*, 4(4):472–480, 1991.

[4] M. J. Atallah, G. N. Frederickson, and S. R. Kosaraju. Sorting with efficient use of special-purpose sorters. *Inf. Process. Lett.*, 27(1):13–15, 1988.

[5] R. Beigel and J. Gill. Sorting $n$ objects with a $k$-sorter. *IEEE Transactions on Computers*, 39:714–716, 1990.

[6] A. Ben-David, N. Nisan, and B. Pinkas. FairplayMP: A system for secure multi-party computation. In *CCS '08: Proceedings of the 15th ACM conference on Computer and communications security*, pages 257–266, New York, NY, USA, 2008. ACM.

[7] D. T. Blackston and A. Ranade. Snakesort: A family of simple optimal randomized sorting algorithms. In *ICPP '93: Proceedings of the 1993 International Conference on Parallel Processing*, pages 201–204, Washington, DC, USA, 1993. IEEE Computer Society.

[8] M. Braverman and E. Mossel. Noisy sorting without resampling. In *SODA '08: Proceedings of the 19th ACM-SIAM Symposium on Discrete algorithms*, pages 268–276, Philadelphia, PA, USA, 2008. Society for Industrial and Applied Mathematics.

[9] B. Brejová. Analyzing variants of Shellsort. *Information Processing Letters*, 79(5):223 – 227, 2001.

[10] R. Canetti, Y. Lindell, R. Ostrovsky, and A. Sahai. Universally composable two-party and multi-party secure computation. In *STOC '02: Proceedings of the thiry-fourth annual ACM symposium on Theory of computing*, pages 494–503, New York, NY, USA, 2002. ACM.

[11] R. Cole. Parallel merge sort. *SIAM J. Comput.*, 17(4):770–785, 1988.

[12] T. H. Cormen, C. E. Leiserson, R. L. Rivest, and C. Stein. *Introduction to Algorithms*. MIT Press, Cambridge, MA, 2nd edition, 2001.

[13] R. Cypher. A lower bound on the size of Shellsort sorting networks. *SIAM J. Comput.*, 22(1):62–71, 1993.

[14] W. Dobosiewicz. An efficient variation of bubble sort. *Inf. Process. Lett.*, 11(1):5–6, 1980.

[15] W. Du and M. J. Atallah. Secure multi-party computation problems and their applications: a review and open problems. In *NSPW '01: Proceedings of the 2001 workshop on New security paradigms*, pages 13–22, New York, NY, USA, 2001. ACM.

[16] W. Du and Z. Zhan. A practical approach to solve secure multi-party computation problems. In *NSPW '02: Proceedings of the 2002 workshop on New security paradigms*, pages 127–135, New York, NY, USA, 2002. ACM.

[17] U. Feige, P. Raghavan, D. Peleg, and E. Upfal. Computing with noisy information. *SIAM J. Comput.*, 23(5):1001–1018, 1994.

[18] G. Franceschini, S. Muthukrishnan, and M. Pătraşcu. Radix sorting with no extra space. In *European Symposium on Algorithms (ESA)*, pages 194–205, 2007.

[19] M. T. Goodrich. The mastermind attack on genomic data. In *IEEE Symposium on Security and Privacy*, pages 204–218. IEEE Press, 2009.

[20] M. T. Goodrich and S. R. Kosaraju. Sorting on a parallel pointer machine with applications to set expression evaluation. *J. ACM*, 43(2):331–361, 1996.

[21] M. T. Goodrich and R. Tamassia. *Algorithm Design: Foundations, Analysis, and Internet Examples.* John Wiley & Sons, New York, NY, 2002.

[22] W. Hoeffding. Probability inequalities for sums of bounded random variables. *Journal of the American Statistical Association*, 58(301):13–30, Mar. 1963.

[23] J. Incerpi and R. Sedgewick. Improved upper bounds on Shellsort. *J. Comput. Syst. Sci.*, 31(2):210–224, 1985.

[24] J. Incerpi and R. Sedgewick. Practical variations of Shellsort. *Inf. Process. Lett.*, 26(1):37–43, 1987.

[25] T. Jiang, M. Li, and P. Vitányi. A lower bound on the average-case complexity of Shellsort. *J. ACM*, 47(5):905–911, 2000.

[26] D. E. Knuth. *Sorting and Searching*, volume 3 of *The Art of Computer Programming*. Addison-Wesley, Reading, MA, 1973.

[27] F. T. Leighton. *Introduction to Parallel Algorithms and Architectures: Arrays, Trees, Hypercubes.* Morgan-Kaufmann, San Mateo, CA, 1992.

[28] T. Leighton. Tight bounds on the complexity of parallel sorting. *IEEE Trans. Comput.*, 34(4):344–354, 1985.

[29] T. Leighton and C. G. Plaxton. Hypercubic sorting networks. *SIAM J. Comput.*, 27(1):1–47, 1998.

[30] B. M. Maggs and B. Vöcking. Improved routing and sorting on multibutterflies. *Algorithmica*, 28(4):438–437, 2000.

[31] D. Malkhi, N. Nisan, B. Pinkas, and Y. Sella. Fairplay—a secure two-party computation system. In *SSYM'04: Proceedings of the 13th conference on USENIX Security Symposium*, pages 20–20, Berkeley, CA, USA, 2004. USENIX Association.

[32] U. Maurer. Secure multi-party computation made simple. *Discrete Appl. Math.*, 154(2):370–381, 2006.

[33] C. McGeoch, P. Sanders, R. Fleischer, P. R. Cohen, and D. Precup. Using finite experiments to study asymptotic performance. In *Experimental algorithmics: from algorithm design to robust and efficient software*, pages 93–126, New York, NY, USA, 2002. Springer-Verlag New York, Inc.

[34] M. Mitzenmacher and E. Upfal. *Probability and Computing: Randomized Algorithms and Probabilistic Analysis.* Cambridge University Press, New York, NY, USA, 2005.

[35] R. Motwani and P. Raghavan. *Randomized Algorithms.* Cambridge University Press, New York, NY, 1995.

[36] M. Paterson. Improved sorting networks with $O(\log N)$ depth. *Algorithmica*, 5(1):75–92, 1990.

[37] C. G. Plaxton and T. Suel. Lower bounds for Shellsort. *J. Algorithms*, 23(2):221–240, 1997.

[38] V. R. Pratt. *Shellsort and sorting networks.* PhD thesis, Stanford University, Stanford, CA, USA, 1972.

[39] S. Rajasekaran and S. Sen. PDM sorting algorithms that take a small number of passes. In *IPDPS '05: Proceedings of the 19th IEEE International Parallel and Distributed Processing Symposium (IPDPS'05) - Papers*, page 10, Washington, DC, USA, 2005. IEEE Computer Society.

[40] J. H. Reif. An optimal parallel algorithm for integer sorting. In *SFCS '85: Proceedings of the 26th Annual Symposium on Foundations of Computer Science*, pages 496–504, Washington, DC, USA, 1985. IEEE Computer Society.

[41] P. Sanders and R. Fleischer. Asymptotic complexity from experiments? A case study for randomized algorithms. In *WAE '00: Proceedings of the 4th International Workshop on Algorithm Engineering*, pages 135–146, London, UK, 2001. Springer-Verlag.

[42] I. D. Scherson and S. Sen. Parallel sorting in two-dimensional VLSI models of computation. *IEEE Trans. Comput.*, 38(2):238–249, 1989.

[43] R. Sedgewick. *Algorithms in C++*. Addison-Wesley, Reading, MA, 1992.

[44] R. Sedgewick. Analysis of Shellsort and related algorithms. In *ESA '96: Proceedings of the Fourth Annual European Symposium on Algorithms*, pages 1–11, London, UK, 1996. Springer-Verlag.

[45] J. Seiferas. Sorting networks of logarithmic depth, further simplified. *Algorithmica*, 53(3):374–384, 2009.

[46] N. Shavit, E. Upfal, and A. Zemach. A wait-free sorting algorithm. In *ACM Symp. on Principles of Distributed Computing (PODC)*, pages 121–128, 1997.

[47] D. L. Shell. A high-speed sorting procedure. *Commun. ACM*, 2(7):30–32, 1959.

[48] M. A. Weiss and R. Sedgewick. Bad cases for shaker-sort. *Information Processing Letters*, 28(3):133 – 136, 1988.